\documentclass[sigplan,screen,nonacm]{acmart}

\renewcommand\footnotetextcopyrightpermission[1]{} 
\AtBeginDocument{%
  }
\usepackage{hyperref}
\usepackage{graphicx}
\usepackage{multirow}
\usepackage{times}
\usepackage{amsmath,amsthm}

\usepackage{amssymb}
\usepackage{algorithm}
\usepackage{mathrsfs}
\usepackage{booktabs} 
\usepackage{color}
\usepackage{hyperref} 
\usepackage{url}
\usepackage{algpseudocode}
\usepackage{pifont}
\usepackage{bm}
\usepackage{comment}
\usepackage{wasysym}
\usepackage{soul}
\usepackage{CJKutf8}
\usepackage{threeparttable}

\def\1{{\bf{1}}}
\def\0{{\bf{0}}}

\def\b{{\bf b}}
\def\c{{\bf c}}

\def\q{{\bf q}}
\def\r{{\bf r}}

\def\x{{\bf x}}

\def\Ncal{{\mathcal{N}}}

\begin{document}

\title{Melody-Guided Music Generation}


\author{Shaopeng Wei}
\affiliation{%
  \institution{School of Business,\\ Guangxi University}
  \country{China}}
\email{shaopeng.wei@gxu.edu.cn}

\author{Manzhen Wei}
\affiliation{%
  \institution{School of Computing and Artificial Intelligence,\\Southwestern University of Finance and Economics}
  \country{China}
}

\author{Haoyu Wang}
\affiliation{%
 \institution{School of Computing and Artificial Intelligence, \\Southwestern University of Finance and Economics}
  \country{China}
 }
 
\author{$\text{Yu Zhao}^*$}
\affiliation{%
  \institution{School of Computing and Artificial Intelligence, \\Southwestern University of Finance and Economics}
  \country{China}
  }
\email{zhaoyu@swufe.edu.cn}

\author{Gang Kou}
\authornote{Y. Zhao and G. Kou are corresponding authors.}
\affiliation{%
  \institution{School of Business Administration, \\Southwestern University of Finance and Economics}
  \country{China}
  }
\email{kougang@swufe.edu.cn}


\begin{abstract}
We present the \textbf{M}elody-\textbf{G}uided \textbf{M}usic \textbf{G}eneration ($\textbf{MG}^2$) model, a novel approach using melody to guide the text-to-music generation that, despite a simple method and limited resources, achieves excellent performance. Specifically, we first align the text with audio waveforms and their associated melodies using the newly proposed Contrastive Language-Music Pretraining, enabling the learned text representation fused with implicit melody information. Subsequently, we condition the retrieval-augmented diffusion module on both text prompt and retrieved melody. This allows $\text{MG}^2$ 
to generate music that reflects the content of the given text description, meantime keeping the intrinsic harmony under the guidance of explicit melody information. 
We conducted extensive experiments on two public datasets: MusicCaps and MusicBench. Surprisingly, the experimental results demonstrate that the proposed $\text{MG}^2$ model surpasses current open-source text-to-music generation models, achieving this with fewer than 1/3 of the parameters or less than 1/200 of the training data compared to state-of-the-art counterparts. Furthermore, we conducted comprehensive human evaluations involving three types of users and five perspectives, using newly designed questionnaires to explore the potential real-world applications of $\text{MG}^2$
 \footnote{We release the code and dataset for reproducing in GitHub: \url{https://github.com/shaopengw/Awesome-Music-Generation}. What's more, $\text{MG}^2$ is being integrated in the Hugging Face Transformers library (See more  \href{https://github.com/shaopengw/Awesome-Music-Generation/issues/3}{here}).}.


  
\end{abstract}





\maketitle

\section{Introduction}
Music generation has been a popular topic in artificial intelligence for decades \cite{hiller1979experimental,van2010music,choi2016text,ji2023emomusictv,yu2021conditional}, and its significance has grown in recent times due to promising applications in generating personalized background music for short videos on platforms like TikTok, YouTube Shorts, and Meta Reels. Additionally, it serves various other scenarios, including dance, virtual reality, and gaming. 



Recently, the development of diffusion models has transformed many fields, including music generation. The ability to generate audio files that can be played directly from given text descriptions has garnered significant attention. A typical type of approach \cite{liu2023audioldm,ghosal2023text} involves encoding the input text description with a large language model, which is then used as a condition in a diffusion model to generate a latent music representation. Subsequently, a decoding module decodes this latent music representation into playable music. However, music, as a quintessential form of art, requires harmony both in the overall composition and within each fragment. Previous methods relying solely on text descriptions to condition the music generation process ensure semantic accuracy but inherently fail to maintain harmony across all fragments. Consequently, these methods face issues such as monotony, noise in generated fragments, and a lack of fluency.

In this paper, we propose the novel Melody-Guided Music Generation ($\text{MG}^2$) model, which leverages melody as guidance to address prior challenges, achieving improved music generation performance.
 We incorporate melody information both implicitly within the Contrastive Language-Music Pretraining (CLMP) and explicitly in the retrieval-augmented diffusion module. Specifically, we propose aligning music waveforms, text descriptions, and melodies within a unified semantic space, where each melody is represented by symbols that reflect the rhythm and harmony of music. This approach enables us to implicitly leverage the expertise embedded in melody while preserving the semantic integrity of text descriptions. Additionally, the alignment performance across these three modalities significantly surpasses that of text description and music waveform alone, demonstrating that the inclusion of the melody modality strengthens the alignment between the other two modalities rather than hindering it.

After alignment, we encode melodies through the trained CLMP to construct a vector database. Text prompts are then encoded to obtain text representations, which are used to retrieve the most relevant melody representations from the melody vector database using a hierarchical retrieval strategy. These representations, along with the retrieved melody guidance, serve as the conditioning inputs to the retrieval-augmented diffusion module. This retrieval-based generation framework endows $\text{MG}^2$ with inherent flexibility and scalability, allowing for seamless replacement and expansion of the melody vector database. Finally, the generated latent music representation is processed through a decoding module comprising a Variational AutoEncoder \cite{liu2023audioldm} and a Vocoder \cite{kong2020hifi} to produce playable music.


Guided by the melody, the generated music not only captures the semantic content of input text but also mitigates monotony, noise, and disjointed melody, ensuring harmony across and within segments. Extensive experiments demonstrate that the proposed $\text{MG}^2$, trained on MusicCaps and MusicBench—two public datasets totaling approximately 132 hours of music and utilizing 416 million parameters, outperforms state-of-the-art open-source models trained with over 200 times the data or 8 times the parameters.

Moreover, music generation emphasizes the artistry of creation, distinguishing it from typical text or image generation. This artistic quality cannot be adequately assessed using objective metrics alone, necessitating subjective evaluation. However, most previous studies \cite{liu2023audioldm, liu2024audioldm, copet2023simple} have not thoroughly or accurately assessed generated music, often relying on fewer than 10 participants and simplistic evaluation criteria. In contrast, we assess the generated music across dimensions of recognizability, text relevance, satisfaction, quality, and market potential by engaging 124 general users, 18 musicians, and 20 short-video creators, using newly designed, comprehensive questionnaires. The results underscore the potential commercial value of $\text{MG}^2$ in real-world applications.

In conclusion, the contributions of this paper are threefold:

\begin{itemize}
    \item We propose $\text{MG}^2$, a novel model that employs melody to guide the music generation process. To the best of our knowledge, this is the first approach to utilize melody in both implicit and explicit manners for generating playable music.




    \item We propose the CLMP module to align text descriptions, music waveforms, and melodies, leveraging melody guidance implicitly. Subsequently, we introduce a retrieval-augmented diffusion module that explicitly utilizes melody by retrieving the most relevant melody to condition the diffusion process. Together, these approaches ensure that the generated latent music representations effectively preserve both semantic content and harmonious rhythm.





    \item We conduct extensive experiments that demonstrate the superior performance of $\text{MG}^2$ and the effectiveness of its sub-modules on two public music datasets. Additionally, we perform a comprehensive human evaluation, involving three types of users and five perspectives, using newly designed questionnaires.
\end{itemize}

\begin{figure*}[t]    
    \centering
    \includegraphics[width=0.99\linewidth]
    {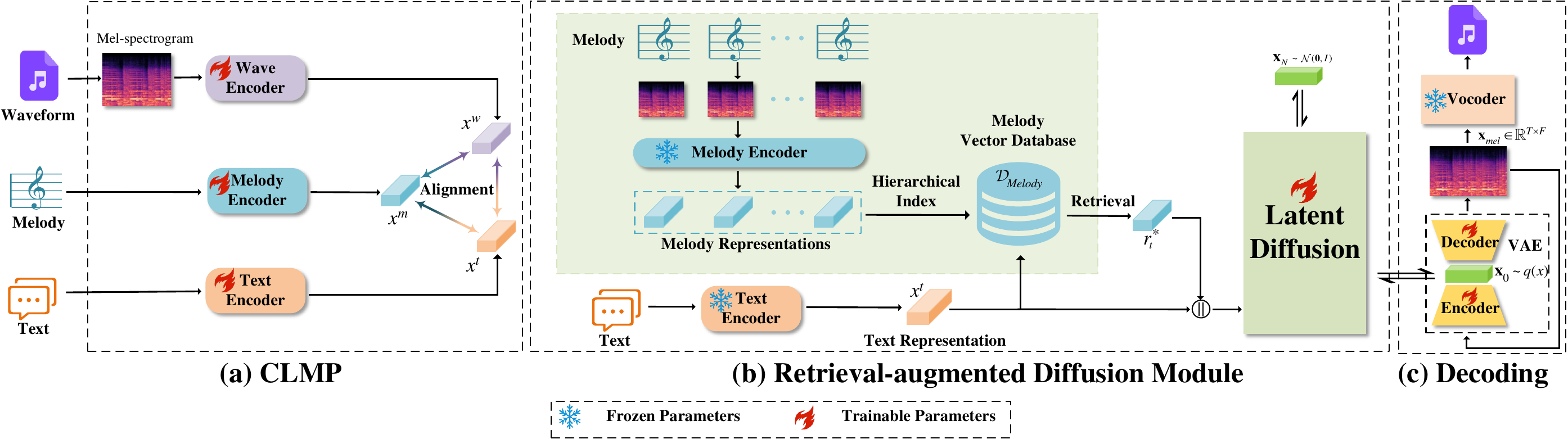}
    \caption{The architecture of the proposed $\text{MG}^2$ comprises three main components: (a) CLMP: This module leverages melody information implicitly by aligning the waveform, melody, and text descriptions within a unified vector space; (b) Retrieval-augmented Diffusion Module: This module generates a music latent vector based on the given text description. It first constructs a melody vector database using previously trained melody representations and retrieves a melody as guidance, and explicitly combines it with the input query to condition the latent diffusion model; (c) Decoding Module: Finally, the decoding module, incorporating a Variational Autoencoder (VAE) and a Vocoder, synthesizes the playable music.}
    \label{fig:framework}
\end{figure*}

\section{Related Works}
\subsection{Text-to-Music Generation}

Text-to-Music (TTM) aims to generate music based on a given text description, forming a specialized sub-domain of Text-to-Audio (TTA) generation \cite{liu2023audioldm,kreuk2022audiogen,yang2023diffsound}. Unlike TTA, TTM requires both semantic alignment with the input description and harmonious melody in the generated music. 
Current research primarily follows two technical approaches: generating $\textit{musical scores}$ \cite{agostinelli2023musiclm,yuan2024chatmusician,ding2024songcomposer} and generating audio files \cite{copet2023simple,chen2024musicldm,lam2024efficient}.
The first approach treats music as a structured language, generating it under a sequence-to-sequence framework. For example, ChatMusician \cite{yuan2024chatmusician} employs ABC notation to produce and interpret symbolic music, while SongComposer \cite{ding2024songcomposer} generates lyrics and melodies in a tuple format, aligning lyrics with musical note attributes.
The second approach aims to produce playable music by directly generating audio files that capture the original sound information. 
For example, AudioLDM \cite{liu2023audioldm} and TANGO \cite{ghosal2023text} use text prompts as guidance to control the audio generation process. To better capture the attributes of music, some works incorporate information on rhythm or melody from various perspectives.
For instance, Music ControlNet \cite{wu2024music} proposed a method that leverages both input text and time-varying attributes—including melody, dynamics, and rhythm—as control signals to guide music generation. Similarly, Mustango \cite{melechovsky2024mustango} utilizes text alongside generated beats and chords to guide the music generation process. Specifically, the beats and chords are derived from the input text using a trained DeBERTa Large model.

However, previous research that uses melody information to control music generation has not aligned melody with text and waveform simultaneously. Consequently, these approaches often fail to preserve both the semantic integrity and the inherent harmony of the generated music.

\subsection{Generative Models}

In recent years, several types of deep neural network-based music generation methods have been developed, including RNN-based models \cite{boulanger2012modeling, waite2016generating}, VAE-based models \cite{roberts2018hierarchical, wang2020pianotree}, GAN-based models \cite{yang2017midinet, donahue2019adversarial} and transformer-based models \cite{donahue2019lakhnes,huang2020pop}. For example, GANSynth \cite{engel2019gansynth} utilized GANs to generate high-fidelity, locally coherent audio.
More recently, diffusion models have become the mainstream approach for music generation. Diffusion models \cite{nichol2021improved, song2020denoising} operate by learning a generative process through a forward step that progressively adds noise to a sample and a reverse step that aims to denoise it. These models have gained significant attention and have been successfully applied to various generation tasks, including image generation \cite{zhang2023adding, saharia2022palette, dhariwal2021diffusion}, text generation \cite{chen2024textdiffuser, wu2023ar, gong2022diffuseq}, video generation \cite{liu2024sora}, and music generation \cite{chen2024musicldm, huang2023noise2music, bai2024seed}.
For instance, AudioLDM \cite{liu2023audioldm} employs CLAP \cite{wu2023large} to align text and audio, using the aligned audio as a condition for the diffusion model to generate latent audio representations during training and text as the condition during inference. 
In contrast,  Seed-Music \cite{bai2024seed} integrates both auto-regressive language modeling with diffusion approaches, enabling the model to handle both music generation and music editing tasks.

However, VAE-based models often produce unrealistic and blurry outputs, GAN-based models are challenging to train and struggle to generate creative works, while RNN and transformer-based models are better suited for generating \textit{music scores} rather than playable music.

\subsection{Retrieval-augmented Generation}

Retrieval-augmented Generation (RAG) enhances the generation process by integrating external knowledge related to the generated content. This approach improves large language models in multiple aspects, such as reducing hallucination \cite{shuster2021retrieval}, safeguarding privacy \cite{zeng2024good}, and enhancing the ability to answer questions with long context \cite{edge2024local}, making it widely adopted. RAG has also been applied in image generation. For instance, Blattmann et al. \cite{blattmann2022retrieval} proposed constructing an independent image database to retrieve a set of nearest neighbors as conditions for the generative model. Similarly, KNN-diffusion \cite{sheynin2022knn} introduced a technique to train a relatively small text-to-image diffusion model without text data, enabling out-of-distribution image generation by swapping the retrieval database during inference.
In the field of audio generation, Yuan et al. \cite{yuan2024retrieval} proposed using retrieved text-audio pairs, given a text prompt as the query, to perform text-to-audio generation.


However, there is no research on applying RAG to enhance music generation.



\section{Preliminary}
\subsection{Retrieval-augmented Generation}
Given the query $\q$, the RAG aims to retrieval the most relevant materials from database $\mathcal{D}_{retrieval}$, thereby enhancing the performance of model. The retrieval process is as following:

\begin{equation}
\begin{array}{l}
\begin{aligned}
    \label{defination-rag}
    \r* = \arg \ \mathop{topK} \limits_{\r \in \mathcal{D}_{retrieval}} Sim(\q,\r),
\end{aligned}
\end{array}
\end{equation}

\noindent where $Sim(\q,\r)$ is the metric of similarity between any piece of sample $\r$ in the database $\mathcal{D}_{retrieval}$ and given query $\q$. $topK$ means that the most relevant $K$ samples will be chosen.


\subsection{Diffusion Model}
Diffusion models \cite{ho2020denoising} have two fundamental processes: (1) forward process injects Gaussian noise into a given sample $\mathbf{x}_0 \sim \boldmath{q}(\mathbf{x})$ with variance schedule $\beta_n \in (0,1)\}_{n=1}^N$ so that the sample $\mathbf{x}_N$ will subject to a standard isotropic Gaussian distribution $\Ncal(\textbf{0},\textbf{\textit{I}})$ after $N$ steps finally; (2) reverse process aims to denoising the damaged sample by predicting the noise $\bm{\epsilon}_n$ in each step $n$ and finally restore the original music.
Specifically, the forward process is as following:

\begin{equation}
\begin{array}{l}
\begin{aligned}
    \label{defination-ldm-forward-1}
    \bm{q}(\mathbf{x}_n| \mathbf{x}_{n-1}) = \Ncal \big(\mathbf{x}_n; \sqrt{1 - \beta_n} \mathbf{n}_{n-1}, \beta_n \bm{I} \big),
\end{aligned}
\end{array}
\end{equation}

\begin{equation}
\begin{array}{l}
\begin{aligned}
    \label{defination-ldm-forward-2}
    \bm{q}(\mathbf{x}_n| \mathbf{x}_{0}) = \Ncal \big(\mathbf{x}_n; \sqrt{\overline{\alpha}_n} \mathbf{x}_{0}, (1-\overline{\alpha}_n) \bm{\epsilon} \big),
\end{aligned}
\end{array}
\end{equation}

\noindent where $\alpha_n = 1 - \beta_n$ and $\overline{\alpha}_n = \prod_{i=1}^n \alpha_i$, which is a reparameterization trick. By using Bayes’ rule and Markov chain assumption, we have following in the reverse process:

\begin{equation}
\begin{array}{l}
\begin{aligned}
    \label{defination-ldm-rerverse-1}
    \bm{q}(\mathbf{x}_{n-1}| \mathbf{x}_{n},\mathbf{x}_{0} ) = \Ncal \big(\mathbf{x}_{n-1}; \Tilde{\mu}_n(\mathbf{x}_n, \mathbf{x}_0),\Tilde{\beta}_n \textbf{I} \big),
\end{aligned}
\end{array}
\end{equation}

\noindent where $ \Tilde{\bm{\mu}}_{n}(\mathbf{x}_{n}, \mathbf{x}_{0})$ and $\tilde{\beta}_{n}$ are defined as follows:

\begin{equation}
 \begin{aligned}
     \Tilde{\bm{\mu}}_{n}(\mathbf{x}_{n}, \mathbf{x}_{0}) &= 
    \frac{\sqrt{\bar{\alpha}_{n}-1} \beta_{n}}{1 - \bar{\alpha}_{n}} \mathbf{x}_{0} + 
    \frac{\sqrt{\bar{\alpha}_{n}(1-\bar{\alpha}_{n}-1)}}{1-\bar{\alpha}_{n}} \mathbf{x}_{n},
    & \\
    \tilde{\beta}_{n} &= \frac{1-\bar{\alpha}_{n-1}}{1-\bar{\alpha}_{n}} \beta_{n}.
\end{aligned}   
\end{equation}

Finally, the training loss for each step is as following:

\begin{equation}
\begin{array}{l}
\begin{aligned}
    \label{defination-ldm-rerverse-2}
    L_n = \mathbb{E}_{n \sim [1, N], \mathbf{x}_0, \boldsymbol{\epsilon}_n} \left[
    ||\boldsymbol{\epsilon}_n - \boldsymbol{\epsilon}_\theta(
        \sqrt{\bar{\alpha}_n} \mathbf{x}_0 + \sqrt{1 - \bar{\alpha}_n} \boldsymbol{\epsilon}_n, n)
    ||^2
\right],
\end{aligned}
\end{array}
\end{equation}

\noindent where $\theta$ is the parameter of  the model that aims to predict the noise in each step.

\section{Method}

As illustrated in Figure \ref{fig:framework}, the proposed method consists of three core components: the CLMP module, the retrieval-augmented diffusion module, and the decoding module. The CLMP is designed to leverage melody guidance implicitly by aligning three modalities: waveform, melody, and text description. Building on this, the retrieval-augmented diffusion module generates latent music representations explicitly guided by the aligned text prompt and the retrieved melody.
Finally, the decoding module transforms this latent representation into playable music.


\subsection{CLMP}
\label{sec-multimodal_alignment}


Unlike CLIP \cite{radford2021learning} and CLAP \cite{wu2023large}, we propose the CLMP, that aligns music waveform, melody, and text description simultaneously. This results in three key alignment subparts: (1) music waveform and melody; (2) melody and text description; (3) music waveform and text description. We refer to CLAP for the implementation of alignment tasks. Specifically, to process the waveform, we first convert each waveform into a mel-spectrogram, which helps to extract the main information from the waveform. We then use HTS-AT \cite{chen2022hts} to encode the mel-spectrogram into audio representations. For the text description, we directly use RoBERTa \cite{liu2019roberta} to encode the input text into text representations.
For both HTS-AT and RoBERTa, we keep the parameters fixed and train two Multilayer Perceptrons (MLPs) following each of these modules, respectively, in a manner similar to prefix-tuning. We also employ a melody encoder to transform the melody into melody representations. However, unlike the waveform and text processing steps, we utilize small, randomly initialized MLP as the melody encoder to process the pitch and duration, rather than a large pretrained model. After encoding, we apply a pooling strategy to convert sequences of varying lengths of melody tokens into a uniform shape, which is then fed into another MLP to generate the updated melody representation.

We then align the three modalities into the same vector space using a contrastive loss function, as follows:

\begin{equation}
\begin{array}{l}
\begin{aligned}
    \label{method-alignment-total_loss}
    L = \frac{1}{6} (L_{mw} + L_{wm} + L_{wt} + L_{tw} + L_{tm} +L_{mt} ),
\end{aligned}
\end{array}
\end{equation}

\noindent where $L_{mw}, L_{wm}, L_{wt} , L_{tw} , L_{tm} , L_{mt}$ denote the contrastive loss between melody and waveform, waveform and melody, waveform and text, text and waveform, text and melody, and melody and text, respectively. Noted that the contrastive loss function is not symmetric.

The calculation of the subpart losses follows a similar structure. For example, the contrastive loss between text and melody is defined as follows:

\begin{equation}
\begin{array}{l}
\begin{aligned}
    \label{method-alignment-tm_loss}
    L_{tm} = \frac{1}{2N} \sum_{i=1}^{N} (\log \frac{\exp(\x_i^t \cdot \x_i^m / \tau)}{\sum_{j=1}^{N} \exp(\x_i^t \cdot \x_j^m / \tau)}),
\end{aligned}
\end{array}
\end{equation}

\noindent where $\tau$ is a learnable temperature parameter. $\x_i^t$, $\x_i^m$ are representations of text and melody, respectively. $N$ is the batch size. We implement contrastive learning within each batch and update the CLMP by batch gradient descent.

\subsection{Retrieval-augmented Diffusion Module}
\label{sec-generation}

\subsubsection{Melody Retrieval}
At this stage, we have achieved alignment of the representations for music, melody, and text using the CLMP. In summary, these representations now exist within a shared semantic space, allowing each modality to retrieve corresponding representations with similar semantic information from the other two modalities. 
First, we employ the CLMP to transform a substantial corpus of melodies into vector representations, constructing the melody vector database $\mathcal{D}_{Melody}$. Next, we obtain the representation of the text prompt, $\x^t$, as a query through the CLMP. To retrieve the most relevant melody $\r^*_t$, we utilize the Hierarchical Navigable Small World Graph (HNSW) \cite{malkov2018efficient}, a hierarchical indexing method, as follows:

\begin{equation}
\begin{array}{l}
\begin{aligned}
    \label{method-retrieval-melody}
  \r^*_t = \mathop{HNSW}\limits_ { \r \in \mathcal{D}_{Melody}} (\x^t,\r).
  
\end{aligned}
\end{array}
\end{equation}

\subsubsection{Conditional Diffusion Process}
We put the retrieved melody $\r^*_t$ as well as input query $\x^t$ together as the conditions of diffusion process as following:

\begin{equation}
\begin{array}{l}
\begin{aligned}
    \label{method-generation-condition}
  \c = W^\top [\x^t || \r^*_t]+ \b,
\end{aligned}
\end{array}
\end{equation}

\noindent where $W^\top \in \mathbb{R}^{2d^\prime \times d^\prime}$ and $\b \in \mathbb{R}^{d^\prime}$ are learnable parameters. $\c$ is the fused condition. 
$d$ and $d^\prime$ denote the dimension of melody or text representation, and dimension of condition $\c$, respectively.

Furthermore, the denoising process with condition $\c$ is as following:

\begin{equation}
\begin{array}{l}
\begin{aligned}
    \label{method-generation-denoising}
    \bm{q}(\mathbf{x}_{n-1}| \mathbf{x}_{n},\mathbf{x}_{0} ) = \Ncal \big(\mathbf{x}_{n-1}; \Tilde{\mu}_n(\mathbf{x}_n, \mathbf{x}_0, \c),\Tilde{\beta}_n \textbf{I} \big),
\end{aligned}
\end{array}
\end{equation}
and the predicted noise is as following:

\begin{equation}
\begin{array}{l}
\begin{aligned}
     \label{method-generation-predicted_noise}
    \overline{\boldsymbol{\epsilon}}_\theta(\mathbf{x}_n,n,\c) = (w+1) \boldsymbol{\epsilon}_\theta(\mathbf{x}_n,n,\c) - w \boldsymbol{\epsilon}_\theta(\mathbf{x}_n,n),
\end{aligned}
\end{array}
\end{equation}

\noindent where $w$ denotes the classifier-free guidance (CFG), which is used to balance the predicted noise with conditional information and that without conditional information.

\subsection{Decoding Module}
\label{sec-decoder}

The decoding module is designed to generate playable music based on the previously generated music latent representation. This module comprises two primary components: a Variational Autoencoder (VAE) and a Vocoder.
The VAE consists of an encoder and a decoder. The VAE encoder transforms the mel-spectrogram \( \x_{mel} \in \mathbb{R}^{T \times F} \) into the latent music representation \( \x_0 \in \mathbb{R}^{C \times \frac{T}{r} \times \frac{F}{r}} \), where \( r \) represents the compression level, \( C \) denotes the number of latent channels, T and F represent the time and frequency dimension size, respectively. Drawing inspiration from the architecture used in AudioLDM, we adopt the same VAE design and training loss functions, which include reconstruction loss, adversarial loss, and a Gaussian constraint loss. During the inference phase, the VAE decoder is utilized to synthesize mel-spectrograms from the music latent representations that have been learned by the retrieval-augmented diffusion module described in Section \ref{sec-generation}.
Subsequently, to convert the mel-spectrogram into playable music, we employ the HiFi-GAN vocoder \cite{kong2020hifi}. This process restores the spectral information into a time-domain signal that can be played as audible music.

\begin{table*}[t]
\caption{Text-to-Music Generation Results. }
\label{tab:main_results}
\begin{threeparttable}
\begin{tabular}{lcccccccc}
\toprule
                 \multicolumn{1}{l}{\multirow{2}{*}{Model}} & \multicolumn{2}{c}{Details} & \multicolumn{3}{c}{MusicCaps}                                                           & \multicolumn{3}{c}{MusicBench}                                                                                                                     \\ \cline{2-9}
                 & Params        & Hours       & FAD↓                     & KL↓                      & IS↑                               & FAD↓                              & KL↓                               & IS↑                           \\ \midrule
MusicLM \cite{agostinelli2023musiclm} & 860M & 280000 & 4.00 &- & -& - &- &-  \\
AudioLDM-M \cite{liu2023audioldm}      & 416M          & 9031        & 3.62                     & 1.78                     & 2.03                              & 2.49                              & 2.21                              & 1.73                                           \\
TANGO  \cite{ghosal2023text}          & 866M          & 145         & 4.05                     & \underline{1.25}                     & -                                 & 1.91                              & \underline{1.19}                  & -                                 \\
MusicGen \cite{copet2023simple} & 3.3B & 20000& 5.55 & 1.53 & 2.07&5.11 & 1.37 & 1.93  \\
AudioLDM 2-Music \cite{liu2024audioldm} & 712M          & 145         & \multicolumn{1}{r}{5.93} & \multicolumn{1}{r}{1.63} & \multicolumn{1}{r}{2.06}          & \multicolumn{1}{r}{4.27}          & \multicolumn{1}{r}{1.51}          & \multicolumn{1}{r}{2.09} \\
AudioLDM 2-Full \cite{liu2024audioldm} & 712M          & 29510       & \multicolumn{1}{r}{\underline{3.21}} & \multicolumn{1}{r}{1.28} & \multicolumn{1}{r}{\underline{2.56}} & \multicolumn{1}{r}{2.19}          & \multicolumn{1}{r}{1.24}          & \multicolumn{1}{r}{2.58}  \\
Mustango  \cite{melechovsky2024mustango}        & 1.4B          & 161         & -                        & -                        & -                                 & \underline{1.74}                              & 1.29                              & \textbf{2.75}                       \\

FluxMusic \cite{fei2024flux} & 2.1B& 22000& 3.52 &1.81 & \textbf{2.74}  & 3.07&1.49 &2.41 \\ 
\midrule
$\textbf{MG}^2$ (Ours)            & 416M          & 132         & \textbf{1.91}            & \textbf{1.21}            & 2.11                              & \multicolumn{1}{r}{\textbf{0.99}} & \multicolumn{1}{r}{\textbf{1.07}} & \multicolumn{1}{r}{\underline{2.62}}                     
\\ \bottomrule
\end{tabular}
\begin{tablenotes}    
        \footnotesize              
        \item[1] The top two scores in each column are marked in bold and underlined.    
      \end{tablenotes} 
\end{threeparttable}
\end{table*}

\section{Experiment}

\subsection{Dataset}

We evaluate all models using the well-known open-access music datasets,  MusicCaps \cite{agostinelli2023musiclm} and MusicBench \cite{melechovsky2024mustango}. 


\textbf{MusicCaps.}
This dataset consists of 5,521 music examples, each of which is accompanied by an aspect list and a text caption written by musicians. The aspect list comprises short sentences such as "high-pitched female vocal melody". This dataset is actually processed based on the AudioSet dataset \cite{gemmeke2017audio}.

\textbf{MusicBench.}
This dataset contains 52,768 music fragments, each with a corresponding text caption. It expands from the MusicCaps dataset. However, we were only able to obtain 42,000 clips of music waveforms from the official link.


Since there are few music datasets that include melody, text, and waveform simultaneously, we construct the melody from the music waveforms.
Specifically, we choose the widely used Musical Instrument Digital Interface (MIDI) as the digital format for melody, which records the start time, duration, pitch, and velocity of each note in a piece of music. However, most music tracks do not have an associated MIDI file. Additionally, current open-source music datasets are primarily collections of music waveforms, with only a small portion including corresponding text descriptions, which are often limited to labels or short phrases. Consequently, finding a dataset that contains waveform files, text descriptions, and MIDI files to align these three modalities is particularly challenging.

We therefore utilize Basic Pitch \cite{Bittner2022Lightweight} to extract melodies from music waveform files and save them as MIDI files. Inspired by Ding et al. \cite{ding2024songcomposer}, 
we retain only pitch and duration information. Each note in the MIDI file is transformed into triplets like \textit{<pitch\_name, duration, rest >}, where duration refers to how long a pitch lasts and rest indicates the time before the next pitch begins. There are 128 types of pitch. Following Ding et al. \cite{ding2024songcomposer}, we divide the continuous duration range up to 6.3 seconds into 512 bins.
An example of a transformed melody might look like: \textit{|<F\#3>,<125>,<79>|<B\#3>,} \textit{<129>,<17>|}, where symbols like \textit{<F\#3>} represent the pitch token, \textit{<129>} is the duration token, and $|$ acts as a separator token.


\subsection{Baselines}
We compare the $\text{MG}^2$ model with the following well-known strong methods, including both text-to-audio models and text-to-music models:

\begin{itemize}
    \item MusicLM \cite{agostinelli2023musiclm}: This work models the conditional music generation as a hierarchical sequence-to-sequence task.
    
    \item AudioLDM \cite{liu2023audioldm}: 
    This model is one of the most notable recent advancements in audio generation, leveraging CLAP, a latent diffusion model, a VAE, and a Vocoder to facilitate text-to-audio generation.

    \item  TANGO \cite{ghosal2023text}: This work employs an instruction-tuned large language model as the text encoder and utilizes a diffusion model for text-to-audio generation, rather than using CLAP for text-audio alignment.

    \item MusicGen \cite{copet2023simple}: This work proposes to use a single-stage transformer LM together with codebook interleaving patterns for conditional music generation.

    \item AudioLDM 2 \cite{liu2024audioldm}: This work proposes converting any audio into a general representation using AudioMAE, and then generating audio, music, or speech through a latent diffusion model, VAE, and Vocoder.

    \item  Mustango \cite{melechovsky2024mustango}: This work introduces a Music-Domain-Knowledge-Informed UNet guidance module to direct the generation of music, effectively capturing music-specific factors for text-to-music generation.
    
    \item FluxMusic \cite{fei2024flux}: This work proposes a diffusion-based rectified flow transformers for text-to-music generation.

\end{itemize}

\subsection{Experimental Setting}

We implement all models using the official codes for the baseline models.
For the CLMP of $\text{MG}^2$, we choose Adam \cite{kingma2014adam} as the optimizer, setting the learning rate to \(1 \times 10^{-5}\), the batch size to 48, and the number of epochs to 90.
For the retrieval-augmented diffusion module and the decoding module of $\text{MG}^2$, we adopt AdamW \cite{loshchilov2017decoupled} as the optimizer, with a learning rate of \(1 \times 10^{-4}\), a batch size of 96, a denoising step of 100, an unconditional guide weight of 3, a compression level of 4, and a total of 60,000 training steps.
During training, we first use audio as the query to develop the model's ability to generate music, then use text as the query to enhance its ability to understand the semantic information of text input.
The CLMP is trained on one NVIDIA RTX 4090 24GB GPU, while the retrieval-augmented diffusion module is trained on one NVIDIA A800 80GB GPU.
Since there is no official code or model checkpoint for MusicLM, we use the reported results from the MusicLM paper.
We use the officially released checkpoint of AudioLDM-M and fine-tune it on the MusicCaps  and MusicBench datasets, respectively.
For TANGO, we refer to the results tested by Mustango.
The training dataset of AudioLDM 2 includes AudioSet, which incorporates MusicCaps and MusicBench. Therefore, we directly test it without fine-tuning.
For Mustango, since it is fine-tuned on MusicCaps and MusicBench, we also directly test it using the official checkpoint on the same test set as $\text{MG}^2$.
For MusicGen and FluxMusic, we use the official code and corresponding largest checkpoints to perform inferences without fine-tuning, following the procedures described in their respective papers.

\subsection{Evaluation Protocol}
We evaluate all models on the MusicCaps and MusicBench datasets. ALL datasets are divided into training, validation, and testing sets at an 8:1:1 ratio through random selection. 
We use the Fréchet Audio Distance (FAD), Inception Score (IS), and Kullback-Leibler (KL) divergence as evaluation metrics. FAD evaluates the similarity between original music and generated music, utilizing VGGish \cite{hershey2017cnn} as the music classifier. IS measures both the quality and diversity of the generated samples. KL divergence measures the distribution distance between the generated samples and the original samples.

\begin{table*}[h]
\caption{Multimodal Alignment Results}
\label{tab:alignment}
\begin{threeparttable}
\begin{tabular}{cccccccccc}
\toprule
\multicolumn{2}{c}{}        & \multicolumn{4}{c}{MusicCaps}                                         & \multicolumn{4}{c}{MusicBench}                                        \\ \cline{2-10}
\multicolumn{2}{c}{}        & R@1             & R@5             & R@10            & mAP@10          & R@1             & R@5             & R@10            & mAP@10          \\ \midrule
\multirow{2}{*}{CLAP} & W2T & 0.5232          & 0.7755          & 0.8622          & 0.6272          & 0.1863          & 0.4139          & 0.5092          & 0.2820          \\
                      & T2W & 0.5154          & 0.7817          & 0.8591          & 0.6257          & 0.1555          & 0.3869          & 0.4862          & 0.2552          \\ \midrule
\multirow{6}{*}{CLMP}  & W2T & \textbf{0.7414} & \textbf{0.9210} & \textbf{0.9566} & \textbf{0.8204} & \textbf{0.2579} & \textbf{0.4926} & \textbf{0.5691} & \textbf{0.3571} \\
                      & T2W & \textbf{0.7043} & \textbf{0.9226} & \textbf{0.9458} & \textbf{0.7976} & \textbf{0.2314} & \textbf{0.4775} & \textbf{0.5649} & \textbf{0.3380} \\
                      & W2M & 0.7631          & 0.9226          & 0.9504          & 0.8347          & 0.6573          & 0.9077          & 0.9434          & 0.7659          \\
                      & M2W & 0.7275          & 0.9211          & 0.9643          & 0.8141          & 0.6329          & 0.8974          & 0.9394          & 0.7481          \\
                      & T2M & 0.5356          & 0.7647          & 0.8359          & 0.6397          & 0.0680          & 0.1769          & 0.2466          & 0.1124          \\
                      & M2T & 0.5309          & 0.7708          & 0.8281          & 0.6311          & 0.0713          & 0.1856          & 0.2553          & 0.1228          \\ \bottomrule
\end{tabular}
    \begin{tablenotes}    
        \footnotesize              
        \item[1] W, T and M denotes Waveform, Text and Melody, respectively.    
      \end{tablenotes} 
\end{threeparttable}
\end{table*}

\begin{figure*}[h]
    \centering
    \includegraphics[width=0.59\linewidth]{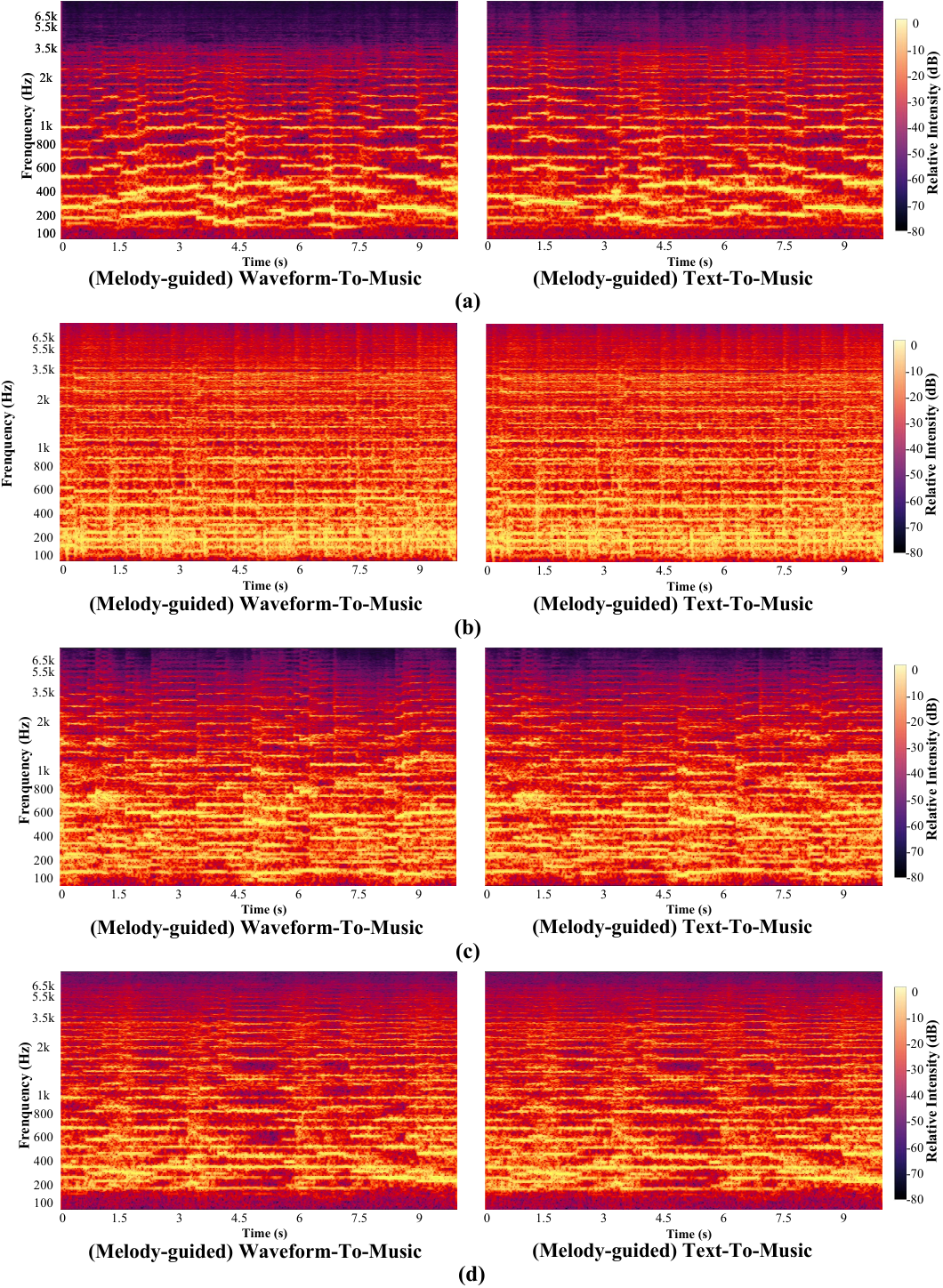}
    \caption{Illustration of accurate alignment.}
    \label{fig:case_study_alignment}
\end{figure*}

\subsection{Music Generation}

\textbf{Main Results.}
As shown in Table  \ref{tab:main_results}, the proposed $\text{MG}^2$ outperforms all baselines in terms of FAD and KL divergence. Specifically, $\text{MG}^2$ exceeds the performance of the strongest baseline, AudioLDM-Full, by 1.30 on MusicCaps and by 0.75 on MusicBench based on the FAD indicator.
Notably, $\text{MG}^2$ is trained on MusicCaps with 5,000 music tracks and MusicBench with 50,000 tracks, totaling approximately 132 hours of audio. In contrast, AudioLDM-Full is trained on 29,510 hours of samples—about 200 times more than the data used by $\text{MG}^2$.
Furthermore, $\text{MG}^2$ achieves these results with only 1/3 of the parameters compared to the strong baseline, Mustango.
Compared to other baselines such as MusicLM, MusicGen and FluxMusic, the gap in training resources and model performance grows larger. Specifically,  $\text{MG}^2$ achieves better performance in terms of FAD and KL indicators using only 1/8 of the parameters of MusicGen, 1/5 of the parameters of FluxMusic, and less than 1/2000 of the training data used by MusicLM.
Additionally, due to the overlap in fine-tuning datasets (i.e., MusicCaps and AudioCaps), models like TANGO and AudioLDM may have been exposed to portions of the MusicCaps test set in our experimental setup. 
Despite this, $\text{MG}^2$ still achieves comparable performance on the IS indicator, further validating its effectiveness. 

\textbf{Analysis.}
Basically, these models can be divided into two types based on the generation module: transformer-based models, including MusicLM and MusicGen, and diffusion-based models, which include the remaining models. Compared to diffusion-based models, transformer-based models require more training data to achieve comparable or better performance. This is likely because transformer-based models need large amounts of data to develop emergent abilities, similar to language models. While transformer-based models excel at modeling temporal information, they struggle with capturing fine-grained patterns, as music is inherently continuous.
AudioLDM, TANGO, AudioLDM 2, and FluxMusic follow a similar approach by using text representations to condition the diffusion process. This ensures that the semantic information of the generated music is preserved, and this capability can be further enhanced with more training data, as demonstrated by comparisons between AudioLDM 2-Music and AudioLDM 2-Full, as well as FluxMusic. Nonetheless, these models still struggle to generate harmonious rhythm.
As a result, melody has played a crucial role in music generation, as evidenced by MusicLM, MusicGen, and Mustango. While both MusicLM and MusicGen use melody solely as an input to guide the generation process, they lack explicit melody guidance for text inputs. Mustango addresses this by predicting melody information from the text input. However, this approach does not provide additional information theoretically, as the amount of input data remains unchanged.
In contrast, $\text{MG}^2$ outperforms all these baselines overall, using fewer model parameters and less training data. Specifically, the melody vector database, serving as an external knowledge base, combined with the diffusion module, makes $\text{MG}^2$ highly efficient, relying on fewer parameters and less training data. The implicit and explicit use of melody guidance ensures that the music generated by $\text{MG}^2$ is harmonious both across fragments and within individual segments, resulting in reduced noise and a more fitting rhythm.



\subsection{Multimodal Alignment}

To demonstrate the effectiveness of the implicit use of melody information, we compare the performance of the alignment module with and without melody information (i.e., CLMP versus CLAP).
As shown in Table \ref{tab:alignment}, we fine-tune the CLMP on MusicCaps and MusicBench using the same checkpoint provided by CLAP. We evaluate the alignment performance using Recall (R@1, R@5, and R@10) and Mean Average Precision (mAP). The results clearly show that the proposed CLMP significantly outperforms CLAP on both datasets and across all metrics. For instance, the CLMP achieves an R@1 score of 0.7414 for the alignment of waveform and text, representing a 21.82\% improvement over CLAP.
Moreover, the CLMP also demonstrates exceptional performance in aligning other modalities. This improvement is likely due to the inherent connection between melody and waveform. Both modalities reflect the patterns of music and thus can enhance each other. This phenomenon is analogous to the real-world scenario in which a student can achieve better grades by simultaneously learning chemistry, physics, and mathematics compared to learning chemistry and physics alone. In this analogy, mathematics plays a crucial role in enhancing the understanding of physics and chemistry, just as melodies enhance the alignment of waveforms and text in our study.

Additionally, we demonstrate the effectiveness of alignment by using both audio waveforms and text descriptions from the same pairs, randomly chosen from the MusicBench dataset, to guide the music generation process. As illustrated in Figure \ref{fig:case_study_alignment}, the music generated with text input and waveform input results in nearly identical music spectrograms in each sub-figure. Due to space limitations, we provide the text descriptions of these pairs in Appendix \ref{appendix: accurate_alignment}. It is important to note that both the audio and text representations are derived from the CLMP. This alignment is due to the semantic correspondence between the text modality and the waveform modality.
On one hand, the results indicate that the text representation and audio waveform representation from the same pair are highly similar. On the other hand, the retrieved melody guidance is also consistent, reflecting the high similarity between the two input queries. In conclusion, the proposed three-modality alignment achieves excellent performance not only in experimental metrics but also in the real music generation process.

\subsection{Ablation Study}

\begin{figure}[t]
    \centering
    \includegraphics[width=0.99\linewidth]{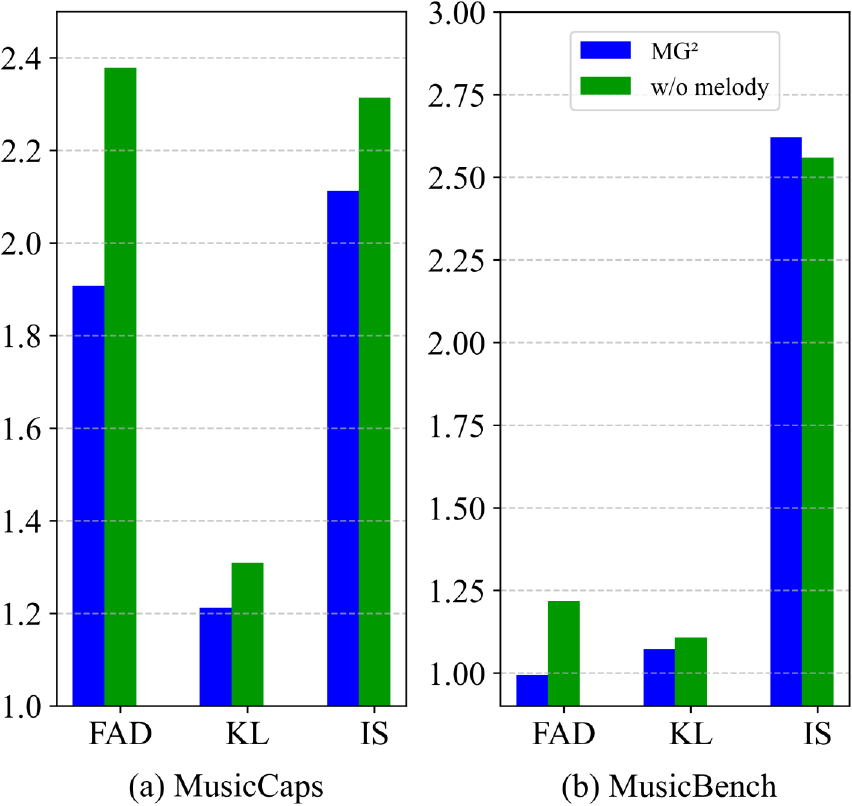} 
    \caption{Ablation Study.}
    \label{fig:ablation_study}
\end{figure}

To demonstrate the effectiveness of the explicit use of melody, we conduct an ablation study on the MusicCaps and MusicBench datasets. As shown in Figure \ref{fig:ablation_study}., we compare the performance of $\text{MG}^2$ with that of $\text{MG}^2$ without melody.
Specifically, we replace the melody component $\r^*_t$ in Equation \ref{method-generation-condition} with zero padding.
The results clearly show that $\text{MG}^2$ outperforms its version without melody, as evidenced by the increase in FAD and KL on both two datasets.
In terms of the IS indicator, the performance improves on the MusicCaps dataset but deteriorates on the MusicBench dataset. In summary, the results
highlight the crucial role that melody plays in enhancing the generation process.

\begin{figure*}[h]
    \centering
    \includegraphics[width=0.6\linewidth]{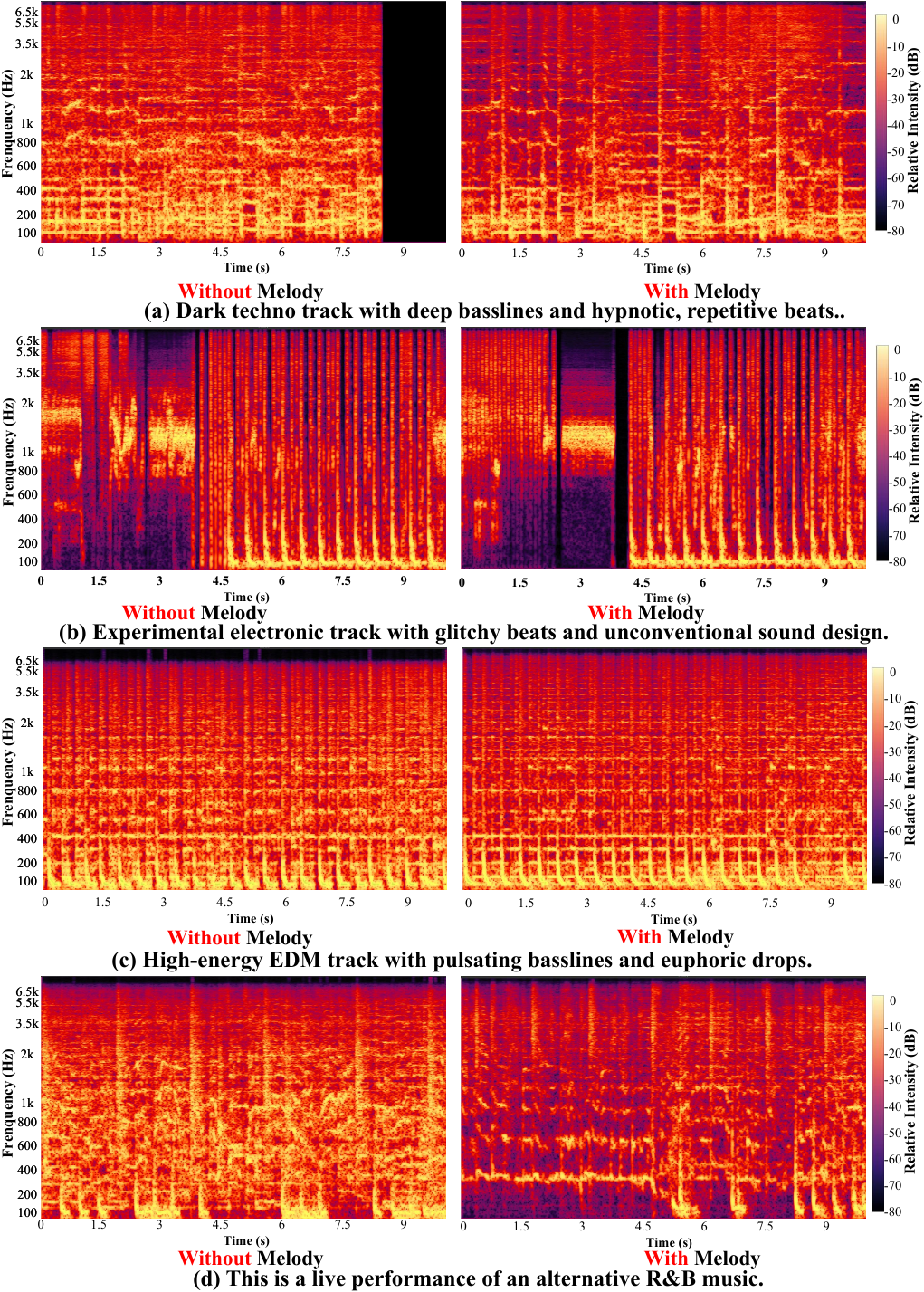}
    \caption{Illustration of melody guidance.}
    \label{fig:case_study_melody}
\end{figure*}

Next, we illustrate the impact of removing melody by presenting four cases in Figure \ref{fig:case_study_melody}.
In Figure \ref{fig:case_study_melody} (a), we observe that deleting the melody information leads to a muted section in the last 2.5 seconds of the generated music. In Figure \ref{fig:case_study_melody} (b), the music generated by $\text{MG}^2$ with melody is clearer and more regular between 1.5 and 4 seconds compared to the version without melody.
In Figure \ref{fig:case_study_melody} (c), removing the melody information results in a loss of semantic clarity. Specifically, both pieces of music are generated using the same prompt emphasizing "High-energy." The version produced by $\text{MG}^2$ displays more high-frequency content than the one generated without melody.
Finally, in the last sub-figure of Figure \ref{fig:case_study_melody}, the spectrogram on the right exhibits a clearer and more evenly distributed frequency and intensity pattern, a characteristic typical of music with a prominent melody. In contrast, the left spectrogram appears somewhat blurred in its spectral details, likely due to the absence of a defined melodic structure.
In conclusion, the inclusion of melody information enables $\text{MG}^2$ to improve the generated music's completeness, clarity, semantic accuracy, and rhythmic coherence.

\begin{figure*}[h]
    \centering
    \includegraphics[width=0.99\linewidth]{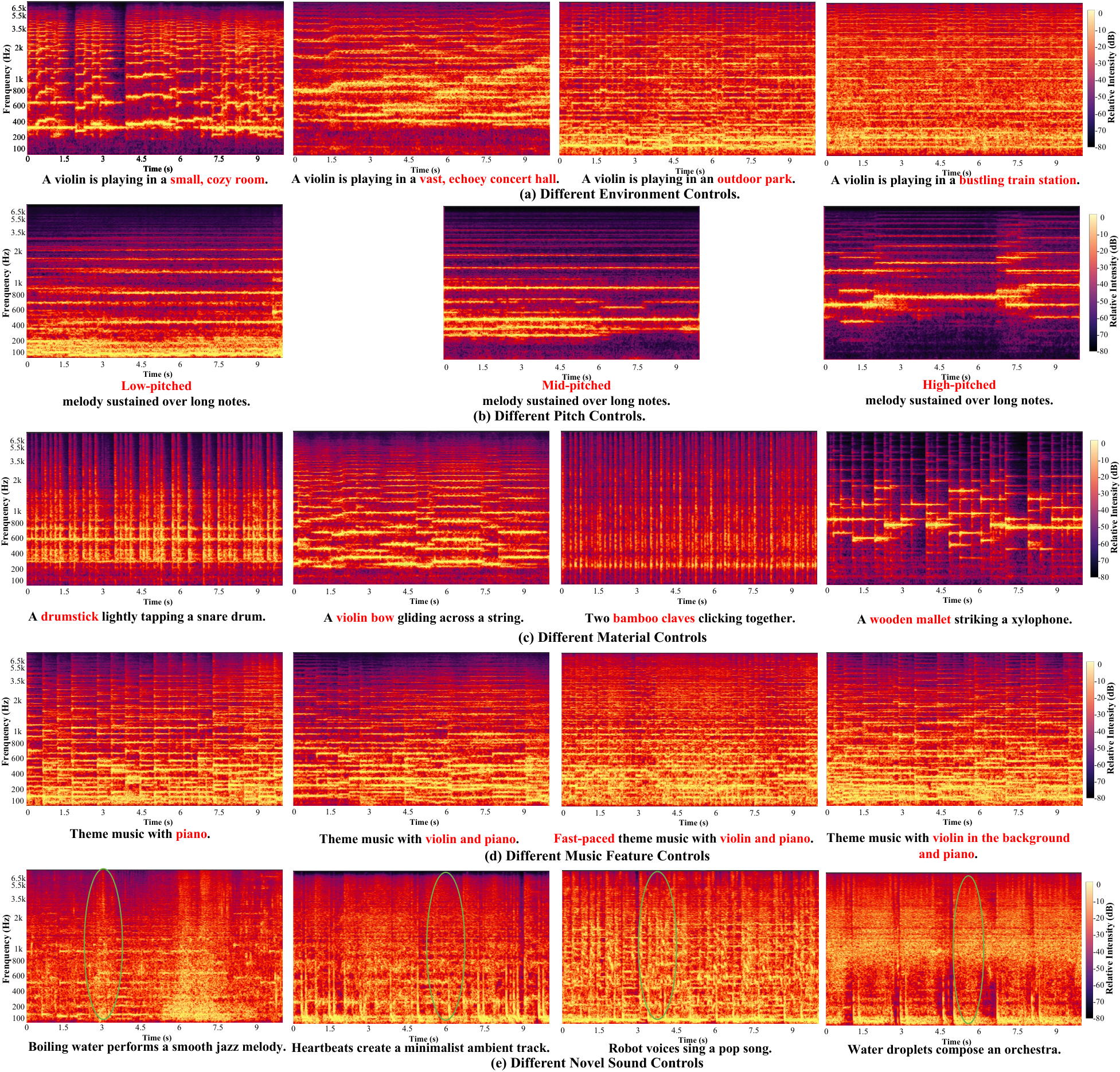} 
    \caption{Illustration of accurate semantic information understanding.}
    \label{fig:semantic_information_control}
\end{figure*}

\subsection{Case Study}

To demonstrate the superiority of $\text{MG}^2$ in understanding semantic information, we present comparisons from various perspectives. In each case, we use three or four prompts that are nearly identical but differ slightly in wording, resulting in subtle changes in the semantic content.
In part (a) of Figure \ref{fig:semantic_information_control}, we observe that $\text{MG}^2$ accurately reflects the environmental context described in the prompts. Specifically, the variations in the prompts correspond to different levels of spaciousness, ranging from a small cozy room to an outdoor park and a bustling train station. The spectrograms reveal similar wave patterns, indicating the use of the same musical instrument—violin—yet they differ in clarity. The violin piece in the small, cozy room is clearer due to the absence of echo, while the violin pieces in the outdoor park and train station have a more distorted, noisy sound, reflecting environmental factors like echo and background chatter, which are evident from the less dark sections in the spectrograms.
In part (b), we observe that $\text{MG}^2$ effectively captures the semantic information related to frequency requirements. For example, the prompt that specifies "High-pitched" results in more high-frequency content compared to the prompt with "Low-pitched," demonstrating the model’s ability to adapt to different frequency characteristics.
In part (c), we show that $\text{MG}^2$ generates music that aligns with the specified musical instrument materials. Different wave patterns in the spectrograms correspond to the different instruments described in the prompts.
Similarly, in part (d), $\text{MG}^2$ generates music that matches the required music instrument, as validated by distinct wave patterns in the spectrograms.
Finally, we highlight $\text{MG}^2$'s ability to generate novel sounds, such as the sound of boiling water, heartbeat, and water droplets, which were not traditionally used in music creation.
We use green circles to label the mel-spectrogram segments that represent novel sounds.
This capability opens up new possibilities for innovative sound generation in video, music, and film production.

\subsection{Parameter Analysis}

\begin{figure}[htp]
    \centering
    \includegraphics[width=0.99\linewidth]{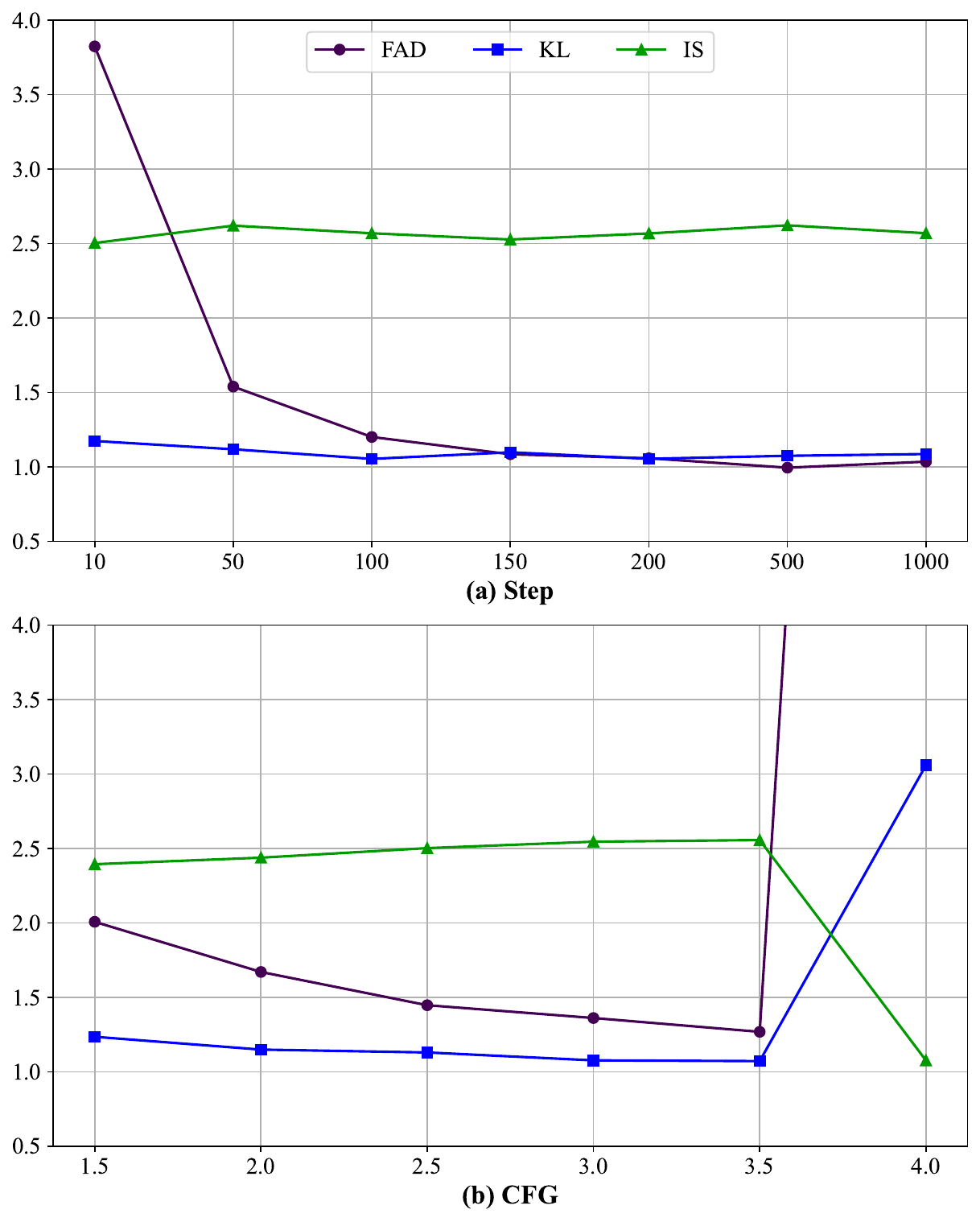}
    \caption{Parameter analysis.}
    \label{fig:parameter_analysis}
\end{figure}

To demonstrate the impact of hyperparameters, we conduct a parameter analysis on two critical settings in the retrieval-augmented diffusion module: the sampling steps of DDIM and the Classifier-Free Guidance (CFG) parameter. Both significantly affect $\text{MG}^2$'s overall performance.
In Figure \ref{fig:parameter_analysis} (a), we observe that as the number of sampling steps increases from 10 to 100, the FAD and KL metrics decrease, while the IS metric improves, indicating an enhancement in $\text{MG}^2$'s performance. Beyond 100 steps, however, further increases yield minimal performance gains, suggesting that while additional steps improve sampling quality, the returns diminish as the step count becomes very large.
Figure \ref{fig:parameter_analysis} (b) illustrates the effects of varying CFG values from 1.5 to 4.0. As CFG increases from 1.5 to 3.5, FAD and KL indicators decrease while the IS indicator improves, demonstrating better overall performance. However, at CFG = 4.0, all metrics worsen, likely due to the CFG mechanism, which balances the predicted error with and without conditioning information in the diffusion's reverse process. While conditional information boosts performance, an excessive CFG value impairs the performance.

\section{Human Evaluation}

We designed questionnaires with demographic questions and specific evaluation criteria for three user groups to comprehensively assess the effectiveness of the proposed $\text{MG}^2$ model. These criteria cover four key aspects: recognizability, text relevance, satisfaction, quality, and market potential (see Appendix \ref{appendix: human_evaluation_measure} for details). Each user group provides unique insights based on their expertise:
General users assess music based on subjective satisfaction, reflecting personal enjoyment but lacking the technical perspective for professional quality evaluation.
Musicians offer professional insights into music quality but may not gauge the business potential.
Short video creators evaluate market potential, leveraging their experience in content creation.
We included seven sample pieces (Table 4) and made them accessible on our \href{https://awesome-mmgen.github.io/}{Website}.

\begin{figure}[h]
    \centering
    \includegraphics[width=0.99\linewidth]{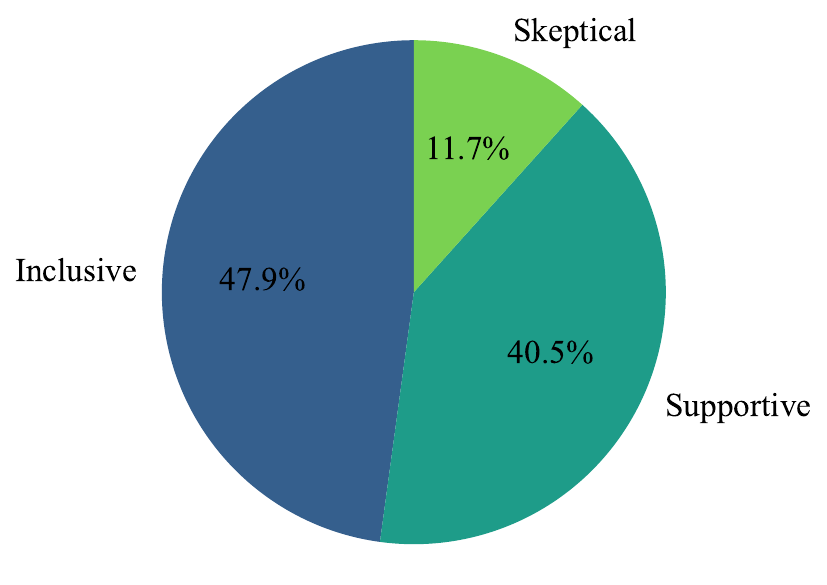}
    \caption{Attitude towards AI music.}
    \label{fig:ai_music_attitude}
\end{figure}

In the demographic section, we also gathered each of the 162 subjects' attitudes toward AI-generated music, displayed in Figure \ref{fig:ai_music_attitude}. Results indicate that 88.4\% of participants are supportive or neutral about AI music, suggesting a promising future for AI in music generation.

\subsection{Recognizability}

\begin{figure}[t]
    \centering
    \includegraphics[width=0.99\linewidth]
    {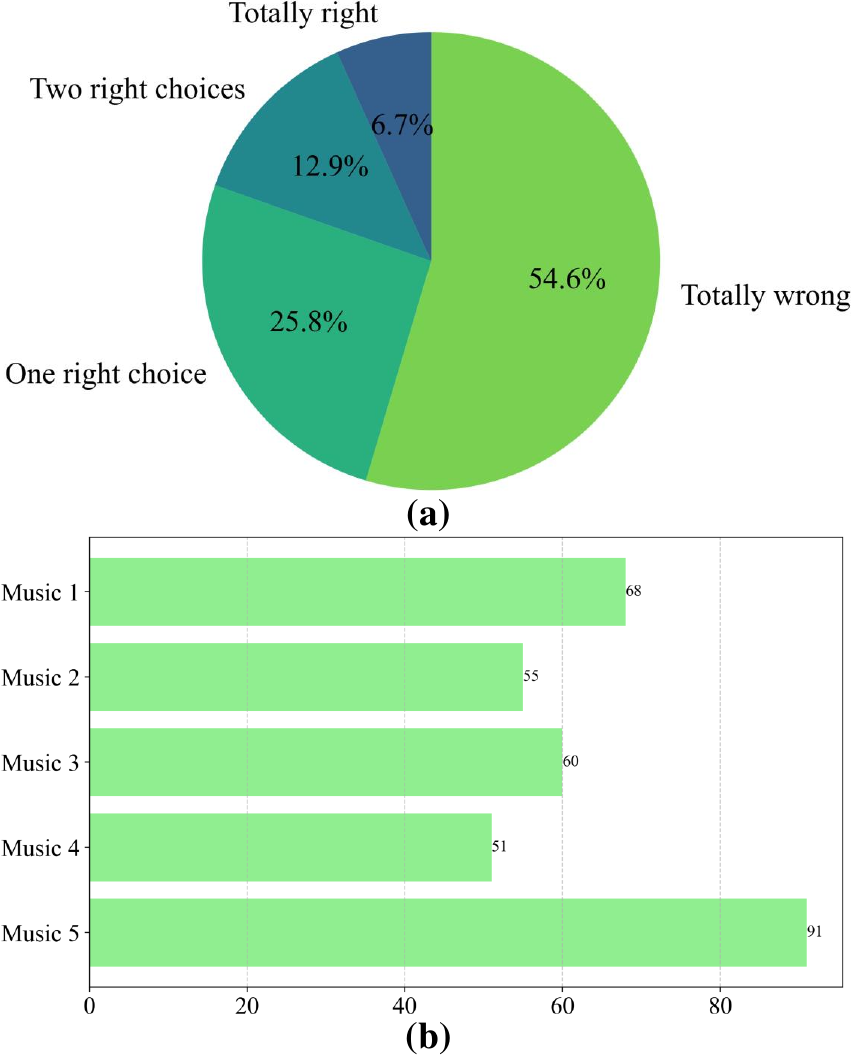}
    \caption{Results of distinguishing AI music.}
    \label{fig:ai_music_analysis}
\end{figure}

We then asked all 162 subjects to identify AI-generated music from a set of five music clips, which included two pieces created by $\text{MG}^2$ and three human-made compositions. Participants could select none, one, or multiple options. Table 3 on our \href{https://awesome-mmgen.github.io/}{Website} provides these music examples for reference.
As illustrated in Figure \ref{fig:ai_music_analysis}, only 6.7\% of participants successfully identified both AI-generated pieces among the five clips. Over half mistakenly labeled all human-made music as AI-generated, failing to identify any of the AI-generated pieces. This result contrasts with AI creations in image and text domains, where people often identify AI content despite high model quality and few visible flaws.
Further analysis of the selections shows that the AI-generated pieces (Music 2 and Music 4) received the fewest “AI music” votes, while human-made tracks, such as Music 5 and Music 1, attracted significantly more. This suggests that $\text{MG}^2$ effectively produces music that many listeners perceive as human-created, underscoring its high quality and realistic output.

\subsection{Text Relevance}

The second part of our questionnaire presents seven pieces generated by $\text{MG}^2$, along with the corresponding prompts and a set of evaluative questions. Each participant hears the prompt and generated music, then responds to questions tailored to their expertise—general users, musicians, and short video bloggers—enabling us to assess the generated music from diverse perspectives. However, all participants are asked to evaluate the alignment between the generated music and its prompt.
Participants select from five options: Very Consistent (VC), Consistent (C), Not Sure (NS), Inconsistent (I), and Very Inconsistent (VI), scored as 5, 4, 3, 2, and 1, respectively, in Table \ref{tab:description relevance}. We calculated the numerical result by averaging responses per music piece, then averaged these scores across all seven pieces to represent $\text{MG}^2$’s prompt relevance score.
As shown in Table \ref{tab:description relevance}, about three-quarters of participants rated the music as either Very Consistent or Consistent with the prompt, indicating a strong alignment. The overall average score of 3.88, close to the 4-point 'Consistent' rating, further supports the high prompt relevance acknowledged by the majority of participants.

\begin{table}[t]
\centering
\caption{Text Relevance}
\label{tab:description relevance}
\resizebox{\linewidth}{!}{
\begin{tabular}{cccccccc}
\toprule
\multirow{2}{*}{\textbf{Music}} & \multicolumn{2}{c}{\textbf{Numerical Results}} & \multicolumn{5}{c}{\textbf{Ratio of different choices /\%}}                       \\ \cline{2-8} 
                                & Average                & Std.                  & VC             & C              & NS             & I              & VI            \\ \midrule
A                               & 3.92                   & 0.88                  & 23.46          & 55.56          & 11.73          & 8.02           & 1.23          \\
B                               & 3.69                   & 1.02                  & 20.37          & 47.53          & 13.58          & 17.28          & 1.23          \\
C                               & 4.02                   & 0.94                  & 30.86          & 52.47          & 6.17           & 8.64           & 1.85          \\
D                               & 3.93                   & 0.90                  & 25.93          & 51.85          & 12.35          & 9.26           & 0.62          \\
E                               & 3.66                   & 1.06                  & 19.14          & 50.00          & 12.35          & 14.81          & 3.70          \\
F                               & 4.11                   & 0.82                  & 32.72          & 52.47          & 8.64           & 5.56           & 0.62          \\
G                               & 3.80                   & 0.96                  & 22.22          & 51.23          & 12.96          & 12.35          & 1.23          \\
\textbf{Final}                & \textbf{3.88}          & \textbf{0.94}         & \textbf{24.96} & \textbf{51.59} & \textbf{11.11} & \textbf{10.85} & \textbf{1.50} \\ \bottomrule
\end{tabular}}
\end{table}

\subsection{Satisfaction}
We conducted a survey with 124 randomly selected participants who do not have specialized knowledge in music. This questionnaire, tailored for general users, included questions about the clarity, infectivity, and freshness of the generated music. More details on the questionnaire can be found in Appendix \ref{appendix_human_evaluation_user}.
The results, presented in Table \ref{tab:user_quetionaire}, show that, on average, 17.71\% and 42.94\% of respondents rated the generated music as 'Very Consistent' and 'Consistent,' respectively, with the desired attributes of clarity, infectivity, and freshness. The overall score of 3.54 further reflects that most participants were generally satisfied with the music generated by $\text{MG}^2$.

\begin{table}[htb]
\centering
\caption{User Satisfaction}
\label{tab:user_quetionaire}
\resizebox{\linewidth}{!}{
\begin{tabular}{cccccccc}
\toprule
\multirow{2}{*}{\textbf{Music}} & \multicolumn{2}{c}{\textbf{Numerical Results}} & \multicolumn{5}{c}{\textbf{Ratio of different choices} /\%}                                         \\ \cline{2-8} 
                                & Average           & Std.              & VC & C     & NS       & I   & VI \\ \midrule
A                               & 3.50              & 1.04              & 13.51           & 46.77          & 18.95          & 17.34          & 3.43              \\
B                               & 3.45              & 1.12              & 18.35           & 37.70          & 18.35          & 22.18          & 3.43              \\
C                               & 3.52              & 1.10              & 18.35           & 42.34          & 16.13          & 19.76          & 3.43              \\
D                               & 3.60              & 1.01              & 17.74           & 44.35          & 19.96          & 16.33          & 1.61              \\
E                               & 3.43              & 1.13              & 18.15           & 37.10          & 18.55          & 22.38          & 3.83              \\
F                               & 3.67              & 0.98              & 18.55           & 46.98          & 18.75          & 14.52          & 1.21              \\
G                               & 3.63              & 1.05              & 19.35           & 45.36          & 16.94          & 15.52          & 2.82              \\
\textbf{Final}                & \textbf{3.54}     & \textbf{1.06}     & \textbf{17.71}  & \textbf{42.94} & \textbf{18.23} & \textbf{18.29} & \textbf{2.82}     \\ \bottomrule
\end{tabular}}
\end{table}

\subsection{Quality}

To assess the quality of the generated music, we gathered responses from 18 musicians with formal music education. These musicians answered professional questions tailored to evaluate the music’s technical and artistic quality. Detailed descriptions of these questions can be found in Appendix \ref{appendix_human_evaluation_musican}.
As shown in Table \ref{tab:musician_quetionaire}, more than half of the musicians, represented by averages of 12.80 and 42.76 in the final row, rated the generated music as high quality. This feedback underscores the professional viability of $\text{MG}^2$’s outputs.

\begin{table}[htb]
\caption{Music Quality}
\label{tab:musician_quetionaire}
\resizebox{\linewidth}{!}{
\begin{tabular}{cccccccc}
\toprule
\multirow{2}{*}{\textbf{Music}} & \multicolumn{2}{c}{\textbf{Numerical Results}}         & \multicolumn{5}{c}{\textbf{Ratio of different choices} /\%}                                                                  \\ \cline{2-8} 
                                & \multicolumn{1}{l}{Average} & \multicolumn{1}{l}{Std.} & \multicolumn{1}{l}{VC} & \multicolumn{1}{l}{C} & \multicolumn{1}{l}{NS} & \multicolumn{1}{l}{I} & \multicolumn{1}{l}{VI} \\ \midrule
A                               & 3.03                        & 1.14                     & 6.25                   & 32.64                 & 20.83                  & 22.92                 & 9.72                   \\
B                               & 3.50                        & 1.03                     & 8.33                   & 50.69                 & 13.89                  & 11.11                 & 5.56                   \\
C                               & 3.54                        & 1.20                     & 20.14                  & 38.89                 & 17.36                  & 11.11                 & 8.33                   \\
D                               & 3.35                        & 1.10                     & 6.94                   & 50.00                 & 13.89                  & 15.97                 & 7.64                   \\
E                               & 2.93                        & 1.11                     & 4.17                   & 32.64                 & 20.14                  & 27.78                 & 9.72                   \\
F                               & 3.47                        & 1.18                     & 14.58                  & 43.75                 & 16.67                  & 9.03                  & 9.72                   \\
G                               & 4.12                        & 0.73                     & 29.17                  & 50.69                 & 11.81                  & 2.78                  & 0.00                   \\
\textbf{Final}                & \textbf{3.42}               & \textbf{1.07}            & \textbf{12.80}         & \textbf{42.76}        & \textbf{16.37}         & \textbf{14.39}        & \textbf{7.24}          \\ \bottomrule
\end{tabular}}
\end{table}

\subsection{Market Potential}

We evaluate the market potential of $\text{MG}^2$ in real creation scenarios by surveying 20 short video bloggers from popular social video platforms in China, including TikTok, REDnote, and Bilibili. These bloggers have fan counts ranging from 1,000 to over 20,000. The questionnaire, detailed in Appendix \ref{appendix_human_evaluation_blogger}, aimed to assess their willingness to use the generated music.
The music used in earlier sections was evaluated on its feasibility for short video editing and the bloggers' willingness to pay for it. The results, presented in Tables \ref{tab:blogger_usefulness} and \ref{tab:up_willingness_to_pay}, show that 75.72\% of bloggers (combining Very Consistent and Consistent responses) believe the music is feasible for short video editing. Additionally, 46.43\% of bloggers expressed a willingness to pay for the generated music.
We also evaluated the usefulness from another perspective by asking bloggers whether they would use the generated music in their content creation. We presented three types of music—exciting, comical, and suspenseful music—each with three clips. Bloggers were asked to select zero to three clips they would use in their creations. As shown in Table \ref{tab:willingness}, over 90\% of bloggers indicated they would use at least one music clip for their content.
These results, reflected in the three tables, demonstrate the practical value of $\text{MG}^2$ from multiple perspectives, highlighting its potential in real-world applications such as video creation.

\begin{table}[htb]
\caption{Usefulness for Short Video Blogger}
\label{tab:blogger_usefulness}
\resizebox{\linewidth}{!}{
\begin{tabular}{cccccccc}
\toprule
\multirow{2}{*}{\textbf{Music}} & \multicolumn{2}{c}{\textbf{Numerical Results}}                        & \multicolumn{5}{c}{\textbf{Ratio of different choices /\%}}                                                                                                                               \\ \cline{2-8} 
                                & Average                           & Std.                              & VC                                 & C                                  & NS                                 & I                                  & VI                                \\ \midrule
A                               & 3.90                              & 0.77                              & 15.00                              & 70.00                              & 5.00                               & 10.00                              & 0.00                              \\
B                               & 3.35                              & 0.85                              & 0.00                               & 60.00                              & 15.00                              & 25.00                              & 0.00                              \\
C                               & 3.90                              & 0.83                              & 20.00                              & 60.00                              & 10.00                              & 10.00                              & 0.00                              \\
D                               & 3.85                              & 0.96                              & 20.00                              & 60.00                              & 10.00                              & 5.00                               & 5.00                              \\
E                               & 3.50                              & 1.12                              & 10.00                              & 60.00                              & 10.00                              & 10.00                              & 10.00                             \\
F                               & 4.30                              & 0.71                              & 45.00                              & 40.00                              & 15.00                              & 0.00                               & 0.00                              \\
G                               & 3.60                              & 1.07                              & 15.00                              & 55.00                              & 10.00                              & 15.00                              & 5.00                              \\
\textbf{Final}                & \multicolumn{1}{c}{\textbf{3.77}} & \multicolumn{1}{c}{\textbf{0.90}} & \multicolumn{1}{c}{\textbf{17.86}} & \multicolumn{1}{c}{\textbf{57.86}} & \multicolumn{1}{c}{\textbf{10.71}} & \multicolumn{1}{c}{\textbf{10.71}} & \multicolumn{1}{c}{\textbf{2.86}} \\ \bottomrule
\end{tabular}}
\end{table}

\begin{table}[htb]
\caption{Willingness to Pay /\%}
\label{tab:up_willingness_to_pay}
\resizebox{\linewidth}{!}{
\begin{tabular}{ccccc}
\toprule
\textbf{Music}   & \textbf{No}    & \textbf{1 Yuan} & \textbf{2-5 Yuan} & \textbf{More than 5 Yuan} \\ \midrule
A                         & 50.00 & 30.00  & 20.00    & 0.00             \\
B                         & 70.00 & 20.00  & 10.00    & 0.00             \\
C                         & 40.00 & 40.00  & 15.00    & 5.00             \\
D                         & 50.00 & 30.00  & 15.00    & 5.00             \\
E                         & 70.00 & 25.00  & 5.00     & 0.00             \\
F                         & 35.00 & 40.00  & 20.00    & 5.00             \\
G                         & 60.00 & 25.00  & 10.00    & 5.00             \\
\textbf{Final}          & \textbf{53.57} & \textbf{30.00}  & \textbf{13.57}    & \textbf{2.86}      \\      \bottomrule
\end{tabular}}
\end{table}

\begin{table}[htb]
\caption{Willingness to use /\%}
\label{tab:willingness}
\resizebox{0.7\linewidth}{!}{
\begin{tabular}{ccccc}
\toprule
                & \multicolumn{1}{l}{0} & \multicolumn{1}{l}{1} & \multicolumn{1}{l}{2} & \multicolumn{1}{l}{3} \\ \midrule 
Part 1           & 15.00                 & 65.00                 & 20.00                 & 0.00                  \\
Part 2           & 0.00                  & 95.00                 & 5.00                  & 0.00                  \\
Part 3           & 15.00                 & 70.00                 & 15.00                 & 0.00                  \\
\textbf{Final} & \textbf{10.00}        & \textbf{76.67}        & \textbf{13.33}        & \textbf{0.00}  \\  \bottomrule  
\end{tabular}}
\end{table}

\section{Conclusion}

In this paper, we propose $\text{MG}^2$, a novel music generation model that leverages melody to guide the generation process both implicitly and explicitly. The model consists of CLMP, retrieval-augmented diffusion, decoding module. 
We conduct extensive experiments to validate the effectiveness of the proposed $\text{MG}^2$ and its subcomponents. Furthermore, we perform a comprehensive human evaluation that demonstrates the feasibility of $\text{MG}^2$ in real-world scenarios.
From an application perspective, future directions include music continuation, music inpainting, long music generation, singing music, and video-to-music generation. Additionally, reducing the computational cost of the music generation model presents a promising avenue for further research.

\section{Acknowledgements}
The research is supported by the Key Technologies Research and Development Program under Grant No. 2020YFC0832702, 
and National Natural Science Foundation of China under Grant Nos. 71910107002, 62376227, 61906159 and Sichuan Science and Technology Program under Grant No. 2023NSFSC0032, and Guanghua Talent Project of Southwestern University of Finance and Economics.





\bibliographystyle{ACM-Reference-Format}
\bibliography{sample-base}

\appendix

\section{Accurate Alignment}
\label{appendix: accurate_alignment}
We present the text descriptions of the used pairs in Figure \ref{fig:case_study_alignment} as follows:

\begin{itemize}
    \item (a) This audio contains someone playing a piece on cello ranging from the low register up into the higher register. This song may be playing during a live performance.
    \item (b) The song is an instrumental. The tempo is medium with a guitar playing a romantic lead, steady drumming, rock drumming, percussive bass line, cymbals crashing and guitar strumming rhythm. The song is exciting and youthful.
    \item (c) This instrumental song features a violin playing the main melody. This is accompanied by a guitar playing chords. The piano plays backing chords. At the end of the song, the piano plays a fill. The bass is played on a double bass. There is no percussion in this song. The mood of this song is romantic. This song can be played in a romantic movie.
    \item (d) This music is instrumental. The tempo is fast with a spirited cello and harpsichord accompaniment. The music is lively, intense, rich, complex , exquisite and engaging. This music a Western Classical.
\end{itemize}

\section{Human Evaluation Measure}
\label{appendix: human_evaluation_measure}

We provide the details of the human evaluation measures used in our study, aiming to establish a standard reference for human evaluation in the music generation field. 

\subsection{Demography Questions}
For all three types of users, we first set the demography section with a multiple-choice question asking users to identify which music clips they believe are AI-generated. The question is as follows:

\begin{itemize}
    \item Do you enjoy listening to music?
    \item Approximately how many hours do you spend listening to music each week?
    \item Can you distinguish among different genres of music?
    \item What types of music do you enjoy?
    \item How old are you?
    \item What is your opinion on AI-generated music?
    \item  Please identify which of the following five music clips you believe are AI-generated (select none, one, or multiple items).
\end{itemize}

\subsection{General User}
\label{appendix_human_evaluation_user}

For general users, the first part of the questionnaire focuses on description relevance, with the following question:

\begin{itemize}
    \item How well does the generated music match the given description?
\end{itemize}

Afterwards, there are 3 questions related to music satisfaction as following:

\begin{itemize}
    \item How would you rate the clarity of the music? Is it free from noise, distortion, or stuttering, and does the melody and rhythm flow naturally?
    \item To what extent does this music evoke emotions (e.g., joy, excitement, relaxation, nostalgia)?
    \item  How fresh and unique does this piece of music feel to you? Does it have a distinct style that stands out from others?
\end{itemize}

\subsection{Musican}
\label{appendix_human_evaluation_musican}

For musicians, we have designed specific questions to evaluate the quality of the generated music from a professional perspective, in addition to sections on demographic information, music relevance, and music satisfaction. The detailed questions are as follows:

\begin{itemize}
    \item How well do you think this piece of music is balanced in terms of high, low, and mid-range tones, contributing to an overall sense of harmony and unity?
    \item How would you assess the rhythm of this piece of music in terms of distinctness and power, with a stable, coordinated beat that provides listeners with a strong sense of motion?
    \item How do you rate the dynamic range of this piece of music, particularly its smooth transition from soft to loud volumes, and its contribution to the expressiveness of the composition?
    \item How would you rate the characteristics of the instruments or sounds in this piece, in terms of clarity, distinctiveness, and individuality of timbres?
    \item How effectively do you think the selection and combination of instruments in this piece complement one another, creating a rich and cohesive musical texture?
    \item  How would you rate the mixing of this piece in terms of spatial depth, with clear layering of tracks that contribute to a vibrant, three-dimensional listening experience?

\end{itemize}

\subsection{Short Video Blogger}
\label{appendix_human_evaluation_blogger}

For short video bloggers, in addition to basic demographic questions and text-description relevance questions, we include two additional types of questions related to the usefulness of the music and their willingness to pay.

\begin{itemize}
    \item To what extent do you believe the music is suitable for integration with visual media (e.g., film, television, games), creating a synergy between sound effects and visual elements?
    \item  Would you be willing to pay for this piece of music?

\end{itemize}

Additionally, we have designed three questions to explore the willingness of bloggers to use music generated by $\text{MG}^2$. Specifically, we selected three music themes, each consisting of three clips, as follows:

\begin{itemize}
    \item Passionate and exciting music, suitable for creations related to sports or esports.
    \item Comical and cheerful music.
    \item Serious and suspenseful music, ideal for documentary, drama, or thriller movie commentary videos.
\end{itemize}

\end{document}